\documentclass{appolb}
\usepackage{epsfig}
\usepackage[english]{babel}
\usepackage{graphicx}
\usepackage{pstricks}
\usepackage{subfigure}
\usepackage{lineno}
\usepackage{placeins}
\usepackage{enumerate}

\makeatletter
\renewcommand{\p@subfigure}{}
\renewcommand{\@thesubfigure}{\alph{subfigure})\hskip\subfiglabelskip}
\makeatother

\begin{document}


\title{
LHC High $\beta^*$ Runs: Transport and Unfolding Methods}

\author{M. Trzebi{\'n}ski$^a$, R. Staszewski$^a$ and J. Chwastowski$^{b,a}$
\address{$^a$Institute of Nuclear Physics PAN,
ul. Radzikowskiego 152,\\
31-342 Krak\'ow, Poland.\\
$^b$Institute of Teleinformatics, \\
Faculty of Physics, Mathematics and Computer Science,\\
                                    Cracow University of Technology,
                ul. Warszawska 24, 31-115 Krak\'ow, \\Poland.
}
}

\maketitle

\begin{abstract}
The paper describes the transport of the elastically and diffractively
scattered protons in the proton-proton interactions at the LHC for the high
$\beta^*$ runs. A parametrisation of the scattered proton transport through the
LHC magnetic lattice is presented. The accuracy of the unfolding of the
kinematic variables of the scattered protons is discussed.
\end{abstract}
\begin{center}
version of: \today
\end{center}

\section{Introduction}

In high energy proton-proton collisions at the LHC most attention is usually
paid to the hard processes. However, soft processes, such as elastic scattering
or diffraction, contribute significantly to the total $pp$ cross section.
Studies of elastically scattered protons are important as this process can be
used to precisely determine luminosity. Measurements of diffracively scattered
protons can contribute to a better understanding of the still not well known
soft QCD.

Scattering angles of protons originating from elastic and diffractive
interactions are very small, of the order of microradians. In order to reach
this angular region in a collider environment, dedicated detectors must be
installed far away (dozens of metres) from the Interaction Point (IP) and very
close to the beam (actually the detectors need to be placed inside the beam
pipe). It is important to point out that typically there are several
accelerator magnets between the IP and such detectors. Therefore, the proton
trajectory depends not only on the scattering angle but also on the proton
energy. It will be shown that from the measurement of the position and
direction of the proton trajectory one can obtain the information about all
components of its momentum after the interaction.

At the LHC there are two sets of detectors that are foreseen to measure
elastically and diffractively scattered protons -- the TOTEM experiment placed
around the CMS Interaction Point and the ALFA detectors of the ATLAS
experiment. This analysis focuses on the ALFA case.

The ALFA experiment aims to determine the absolute luminosity of the LHC at the
ATLAS from the measured rate of elastic scattering events in the
Coulomb-nuclear amplitude interference region \cite{TDR}. It is worth mentioning that the ALFA
detectors offer a possibility to study other processes, \textit{eg.} single diffraction or 
even exclusive production \cite{piony}.
For such measurements
a special tune of the LHC machine is required. Such dedicated optics must
deliver:
\begin{enumerate}[(a)]
  \item a very large value of the betatron function at the IP ($\beta^{*}$),
  \item 90$^o$ phase advance of the betatron function between the IP and
    the detector locations in at least one transverse direction,
  \item a small emittance ($\epsilon^{*}$) of the beams. 
\end{enumerate}
In fact, such optics provides parallel-to-point focusing in the
$(y,z)$ plane. A solution fulfilling the above requirements is
called the high $\beta^{*}$ optics. 

In the first phase (hereafter called the \textit{early high $\beta^{*}$}) the
LHC is foreseen to run with the beam of 3.5 TeV energy, $\beta^{*} = $~90~m and
$\epsilon^{*} = 2.5\ \mu$m$\cdot$rad. In the second phase (the \textit{nominal
high $\beta^{*}$}) the beam energy will be equal to 7 TeV, $\beta^{*}$ to
2625~m and $\epsilon^{*}$ to 1 $\mu$m$\cdot$rad. One should note that the
nominal LHC $\beta^{*}$ value for high luminosity runs is 0.55~m. The machine
special settings for the high $\beta^{*}$ optics are described in
\cite{highbeta, sophie}.

\section{Transport Simulation}

To compute the particle trajectory in a magnetic structure of an accelerator
one of several dedicated transport programs can be used. In this paper MAD-X
\cite{MADX}, a program used to design and simulate particle beam behaviour
within an~accelerator, was employed. The program allows to perform the
calculations using the thick lens approximation of the magnets -- the
Polymorphic Tracking Code (PTC) module \cite{PTC}. This module takes into
account not only the magnetic structure and the geometry of the beam chamber,
but also the fringe fields and edge effects.  It is important to point out
that, contrary to the studies of the beam, the thin lens approximation is valid
for protons of the beam but not for the ones scattered in interactions.  This
is because the latter are more deflected in the magnetic field, hence their
distance from the magnets centres can be large and additional effects can play
an important role.

\begin{figure}[th]
\begin{center}
\includegraphics[width=0.80\columnwidth]{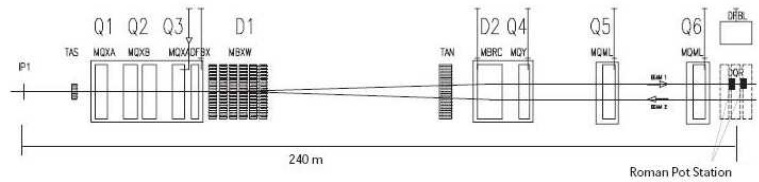}
\caption{The LHC magnet structure close to the ATLAS Interaction Point.
\label{LHC_magnets}}
\end{center}
\end{figure}

The LHC magnetic lattice in vicinity of the ATLAS IP is presented in Fig.
\ref{LHC_magnets}. Quadrupole magnets are labelled with letter $Q$ and dipole
magnets with letter $D$. In the following, a reference frame with the $x$--axis
pointing towards the accelerator centre, the $y$--axis pointing upwards and the
$z$--axis along one of the beams is used. All presented calculations were
performed for the $beam\ 1$ that performs the clockwise motion.  However, the
results are qualitatively relevant also for the $beam\ 2$, which does the
counter clockwise rotation.

The ALFA experiment consists of four detector stations placed symmetrically
with respect to the ATLAS IP at 237.4 m and 241.5 m. In each station there are
two roman pot devices, which allow to insert the position sensitive and
triggering detectors vertically into the beam pipe. Two stations are needed at each
side to be able to measure not only the scattered proton trajectory position, but
also its direction (elevation angles).

\begin{figure}[hbtp]
\begin{center}
\subfigure[early high $\beta^{*}$]{
\includegraphics[width=0.45\columnwidth]{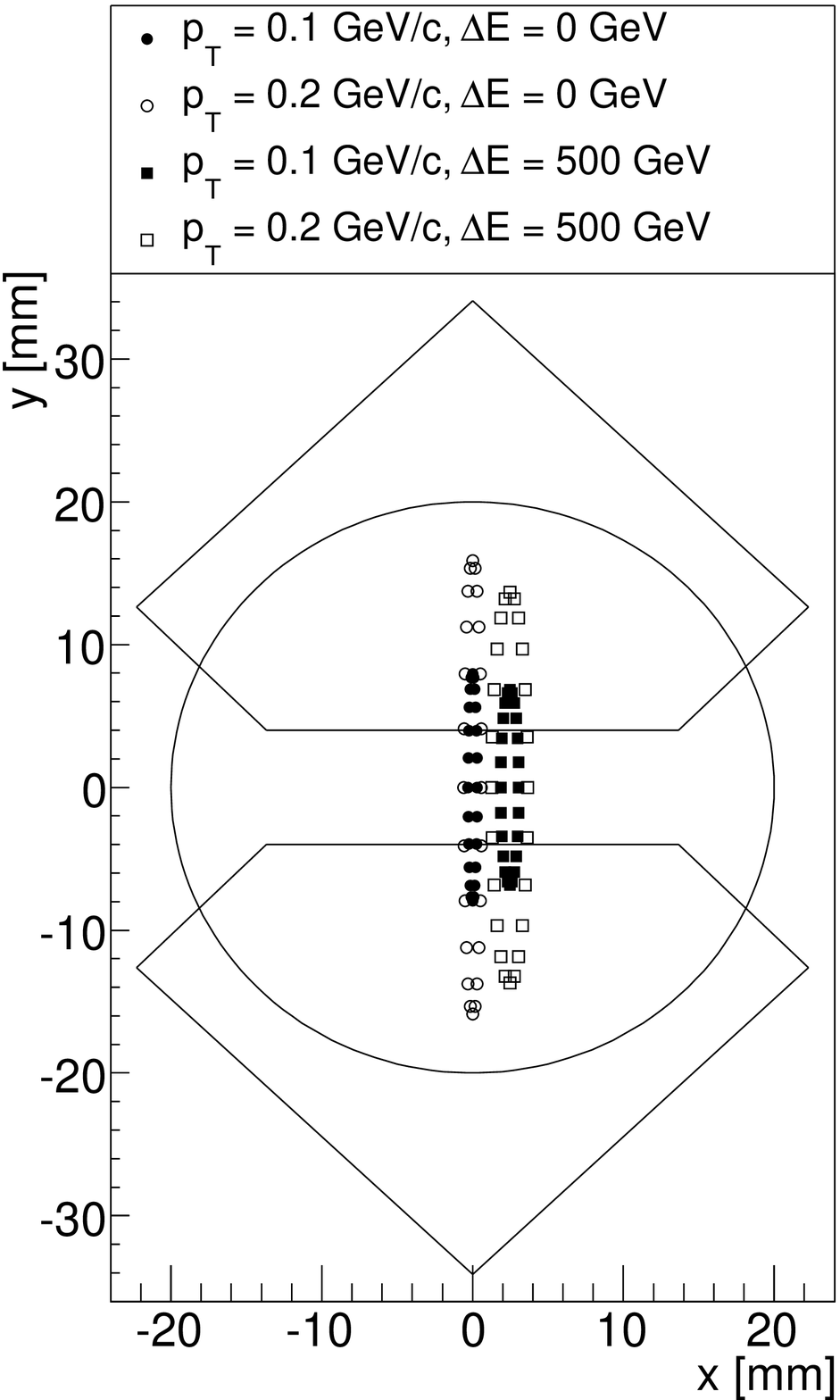}
\label{fig_rings_35}
}
\subfigure[nominal high $\beta^{*}$]{
\includegraphics[width=0.45\columnwidth]{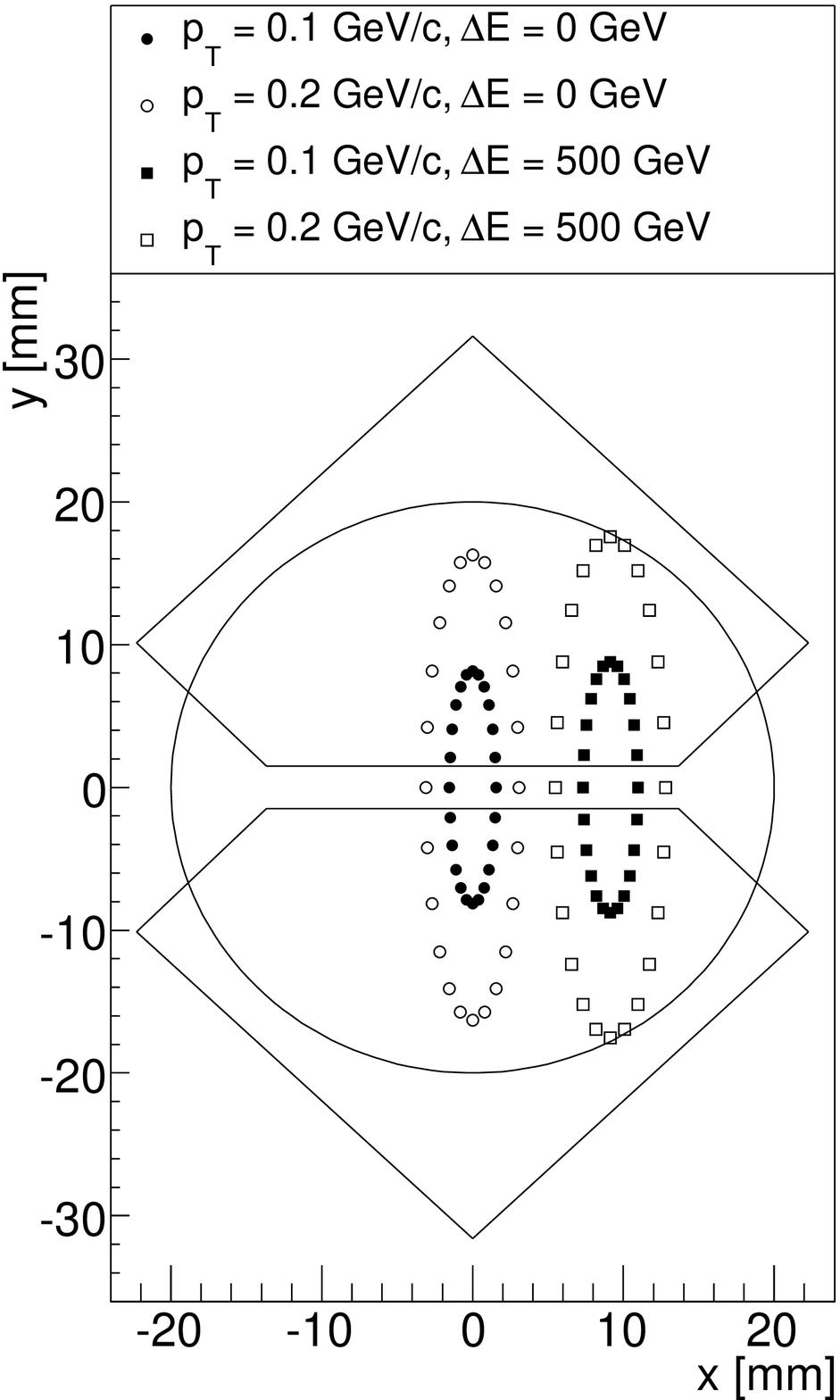}
\label{fig_rings_70}
}
\end{center}
\caption{The positions of protons with different energy loss ($\Delta E$) and
transverse momentum ($p_{T}$) at the first ALFA station for the two different
LHC optics settings. The solid lines mark the beam pipe aperture and the ALFA
detector active area.
\label{fig_rings}}
\end{figure}

An important point is to understand the dependence of the scattered proton trajectory
position in the detector on its energy and momentum. It is illustrated in Figs
\ref{fig_rings_35} and \ref{fig_rings_70} for early and nominal high
$\beta^{*}$ optics, respectively. For both optics settings the impact of the
$p_{y}^{\mathrm{IP}}$-momentum component at the IP on the proton position in
the detector station is much greater than that due to $p_{x}^{\mathrm{IP}}$.
The higher is the proton energy loss, $\Delta E$, the higher is its deflection
towards the machine centre. However, due to the differences in the LHC optics
for both tunes this deflection is larger in the case of the nominal high
$\beta^{*}$.

\begin{figure}[th]
\begin{center}
\subfigure[early high $\beta^{*}$]{
\includegraphics[width=0.45\columnwidth]{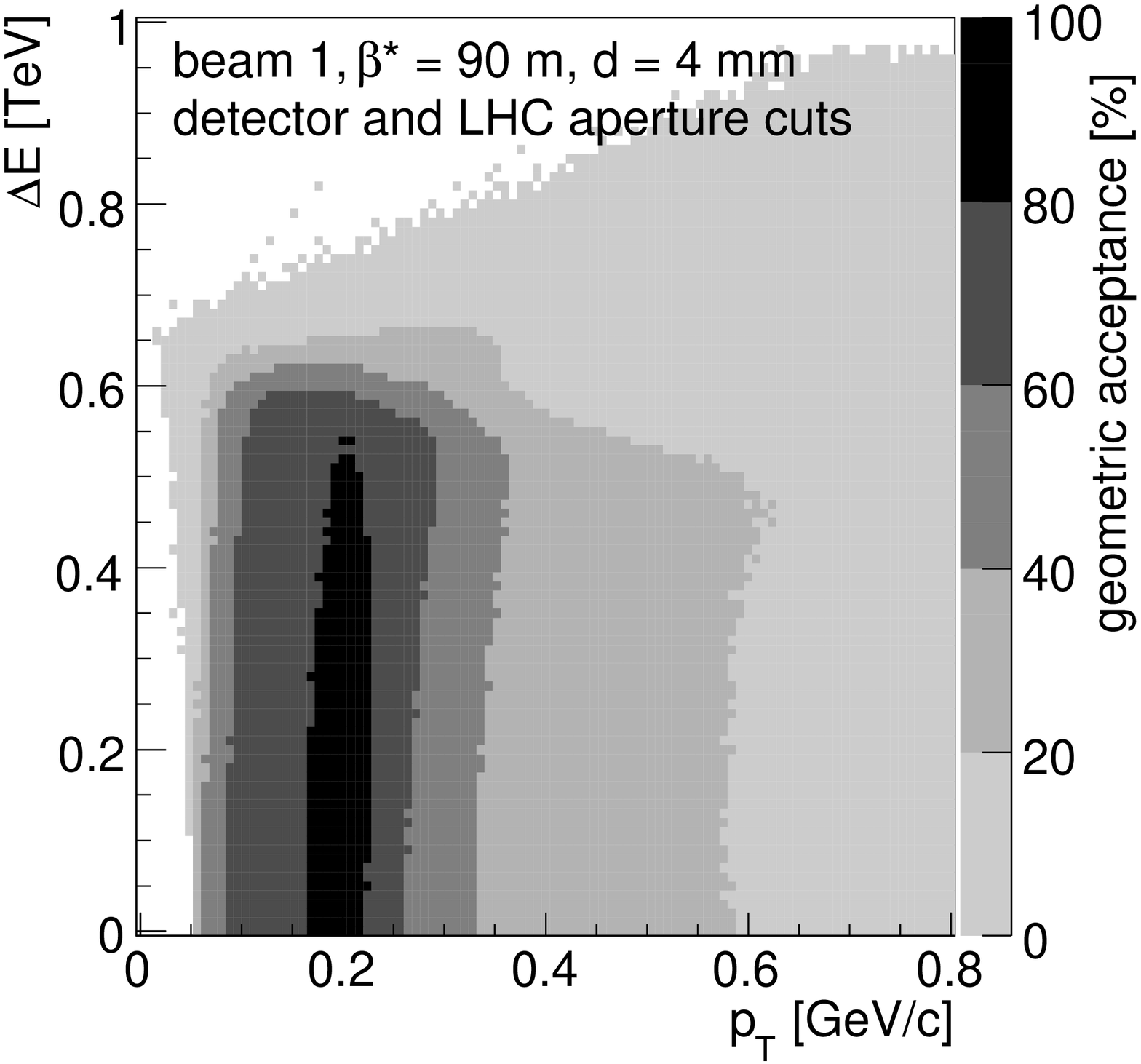}
\label{fig_acceptance_detector_35}
}
\subfigure[nominal high $\beta^{*}$]{
\includegraphics[width=0.45\columnwidth]{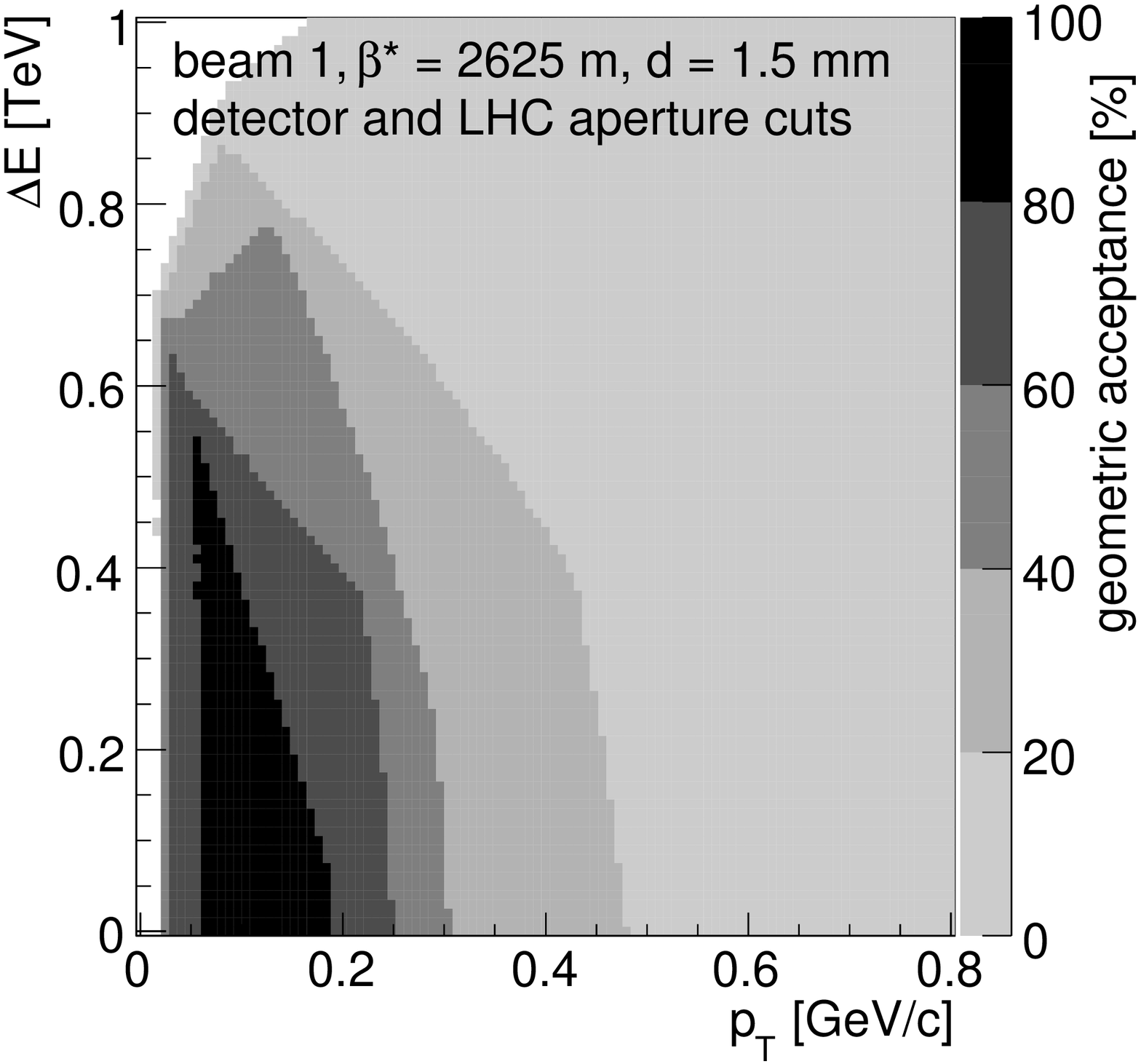}
\label{fig_acceptance_detector_70}
}
\end{center}
\caption{The geometrical acceptance of the ALFA detector as a function of the
proton energy loss ($\Delta E$) and its transverse momentum ($p_T$) for two LHC
settings. The distance between the beam centre and the detector edge was set to
4 mm for early high $\beta^*$ and for 1.5 mm for nominal high $\beta^*$.
\label{fig_acceptance_detector}}
\end{figure}

Naturally, not all scattered protons can be measured in the ALFA
detectors. Such proton can be
too close to the beam to be detected or it can hit the LHC elements (a
collimator, the beam pipe) in front of the ALFA station. The geometric
acceptance, shown in Fig. \ref{fig_acceptance_detector} for both optics settings,  is defined as a ratio of the number of protons of a given energy
loss ($\Delta E$) and transverse momentum ($p_{T}$) that crossed the active
detector area to the total number of the scattered protons having $E$ and
$p_{T}$. In the calculations the following factors were taken into account:
the beam properties at the IP, the beam chamber and the detector geometries,
the distance between the detector edge and the beam centre. This distance was
set  4 and 1.5 mm -- the values expected for the early and nominal high $\beta^{*}$ runs.
Values of other parameters are listed in Table~\ref{tab_smearings}.

\begin{table}[htbp]
	\caption{The LHC beam parameters at the ATLAS IP for early and nominal high~$\beta^{*}$ runs.}
	\label{tab_smearings}	
	\begin{center}	
		\begin{tabular}{c c c c }
	   \hline \\ [-1.5ex]
		Parameter & Unit & Early High $\beta^{*}$ & Nominal High $\beta^{*}$\\ [1ex] 
\hline \\[-1.5ex]
		$\sigma_{x_{\mathrm{IP}}}$, $\sigma_{y_{\mathrm{IP}}}$ & mm & 0.3 & 0.612 \\
		$\sigma_{x_{\mathrm{IP}}'}$, $\sigma_{y_{\mathrm{IP}}'}$ & $\mu$rad & 3.33 & 0.233 \\
		$\sigma_{p_{T}}$ & MeV/c & 11.7 & 1.6 \\ [1ex]
\hline 
		\end{tabular}
	\end{center}
\end{table}

As can be observed, the region of acceptance above 80\% is limited by $\Delta E
< 0.6$~TeV and $ 160$~MeV/c $< p_T <$ 200~MeV/c for early high $\beta^{*}$.  If
the requirement on the acceptance value is lowered to 60\% then the range of
the accepted transverse momentum values gets slightly larger. In the case of
the nominal high $\beta^{*}$ the high acceptance region is a triangle-like and
spans a bit larger range of the proton transverse momentum.  Moreover, it is
worth mentioning that high $\beta^*$ runs will have very low both instantaneous and integrated
luminosities. This implies that only a few events containing particles particles
with high energy loss are expected to be observed. Therefore, the most
important factor is the accepted $p_T$ range for small $\Delta E$.

\begin{figure}[ht]
\begin{center}
\subfigure[early high $\beta^{*}$]{
\includegraphics[width=0.45\columnwidth]{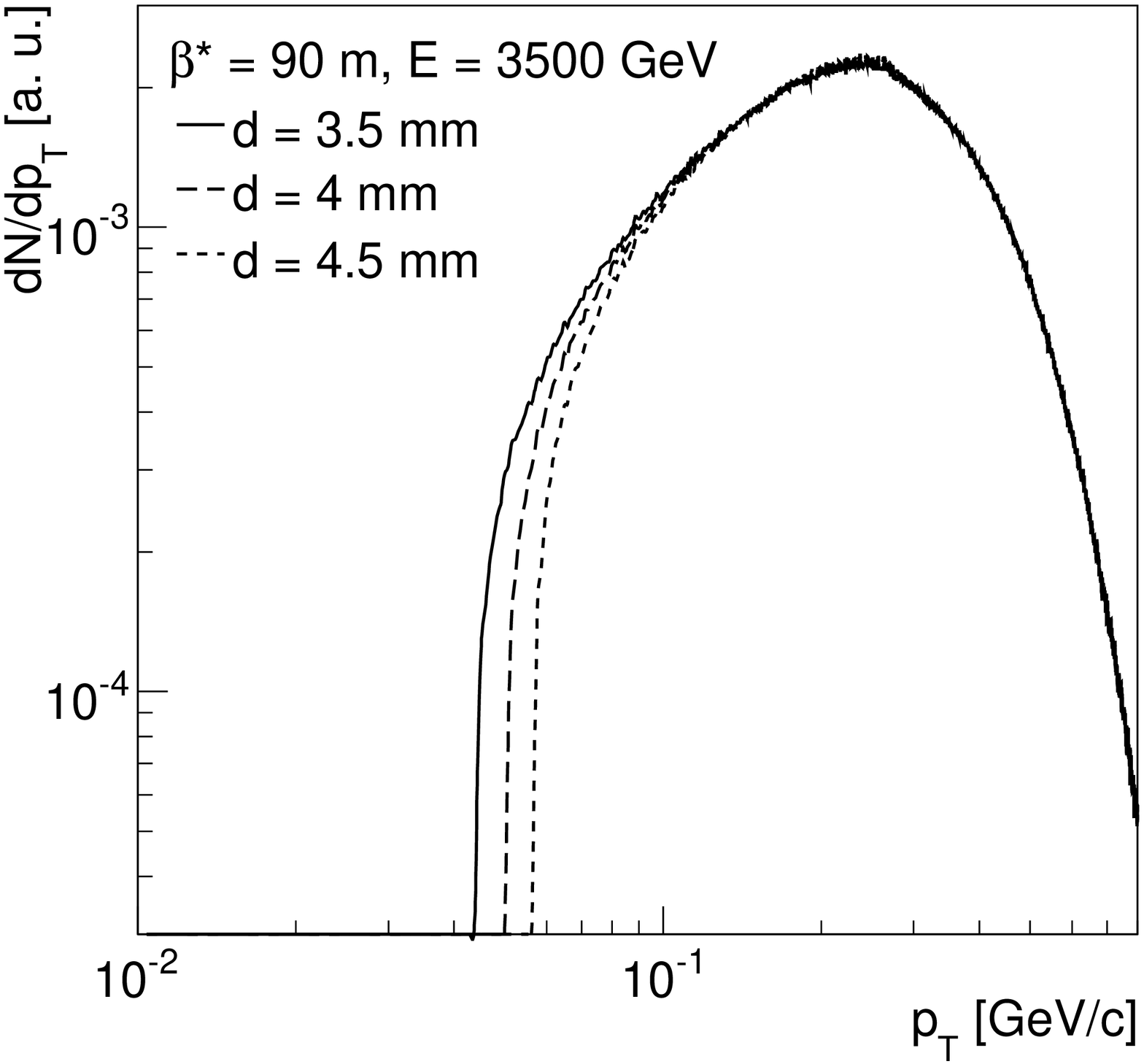}
\label{fig_acceptance_detector_1dim_35}
}
\subfigure[nominal high $\beta^{*}$]{
\includegraphics[width=0.45\columnwidth]{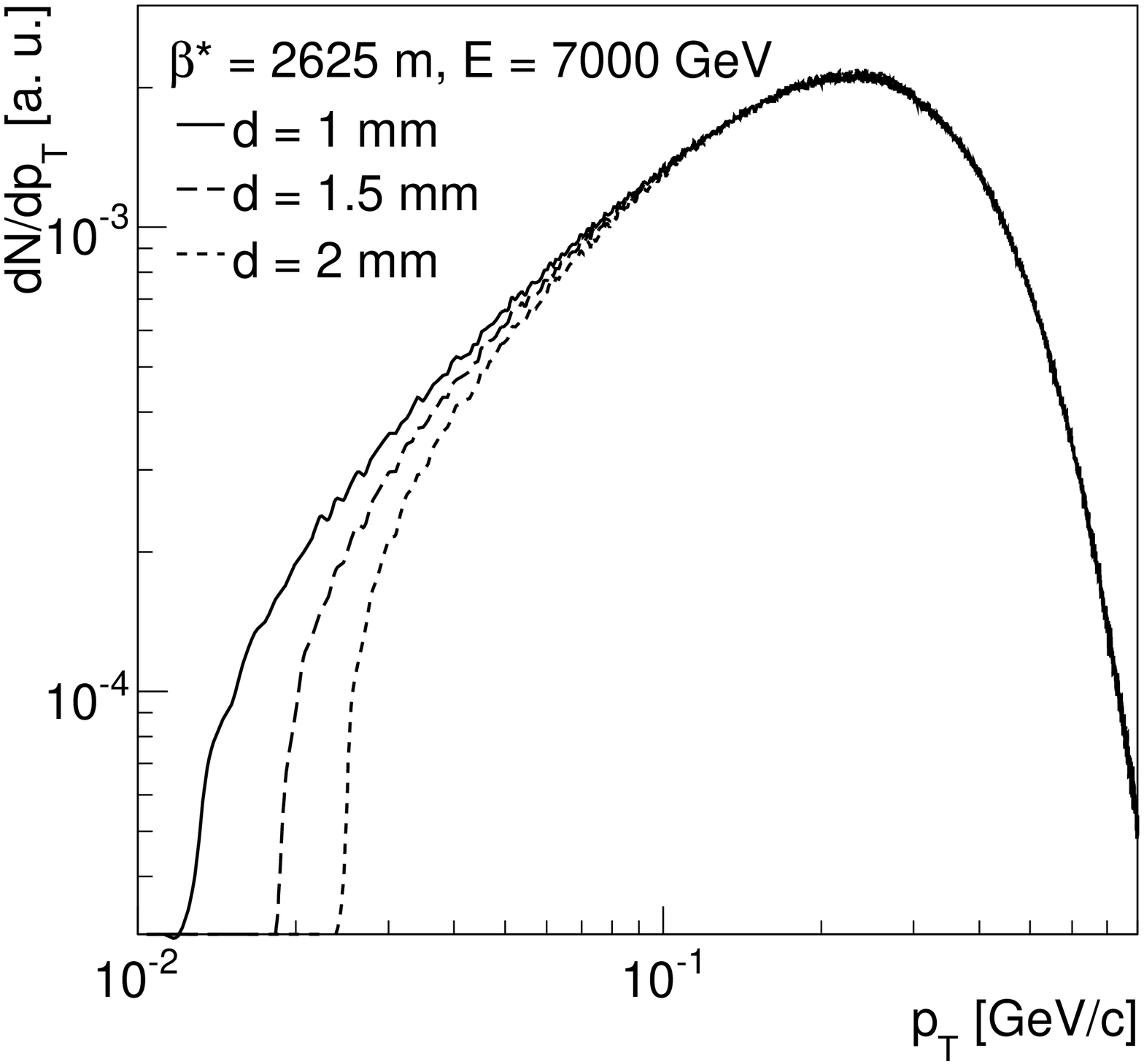}
\label{fig_acceptance_detector_1dim_70}
}
\end{center}
\caption{The influence of the distance between the detector edge and the beam
centre on the accepted proton transverse momentum.
\label{fig_acceptance_detector_1dim}}
\end{figure}

The minimum transverse momentum of protons that can be registered depends on
the distance between the detector edge and the beam centre. This is
demonstrated in Fig.  \ref{fig_acceptance_detector_1dim} where the scattered
proton transverse momentum spectra are shown for different distances and both
machine settings. Clearly, the smaller the distance between the beam and the
detector, the smaller is the limiting value of the accepted proton's $p_T$.
This is particularly important for the elastic scattering measurement where the
possibility of reaching as small $p_T$ values as possible is crucial.

\section{Transport Parametrisation}

Anticipated positions of protons at the detector for a given momentum can be obtained from
the transport simulation. One can also prepare a look-up table and interpolate
the positions. The disadvantage of the above methods is either a long
calculation time as in the first case or very extensive use of the computer storage
(as size of the looking table grows with the required precision) in the second.
The alternative is to use the transport parameterisation, which is very fast
and requires only very little memory space. Moreover, the parameterisation
provides an analytical representation of the proton position and momentum. This
idea was first proposed in \cite{Staszewski}.

The LHC magnetic structure in the vicinity of the ATLAS detector is described
only by the drift spaces, the dipole and the quadrupole magnets (\textit{cf.}
Fig. \ref{LHC_magnets} or \cite{LHC_magnets}). Therefore, a linear transport
approximation can be applied to describe the scattered proton transport.

For a transverse variable ${\zeta} \in \{x, y, x', y'\}$ the transport can be
effectively described by the following equations:
\begin{eqnarray}
  \zeta & = & A_{\zeta} + B_{\zeta}\cdot x_{\mathrm{IP}} + C_{\zeta}\cdot y_{\mathrm{IP}} + D_{\zeta}\cdot z_{\mathrm{IP}} + E_{\zeta}\cdot x_{\mathrm{IP}}' \label{eq:par1}\\
  & & + F_{\zeta}\cdot y_{\mathrm{IP}}' + G_{\zeta}\cdot z_{\mathrm{IP}} \cdot x_{\mathrm{IP}}' + H_{\zeta}\cdot z_{\mathrm{IP}}\cdot y_{\mathrm{IP}}', \nonumber
\end{eqnarray}
where $A_{\zeta}, \ldots, H_{\zeta}$ are the polynomials in the reduced energy
loss ($\xi = \Delta E/E_{beam}$) of rank $k_{A_{\zeta}},\ \ldots,\
k_{H_{\zeta}}$: $$A_{\zeta} = \sum_{n = 0}^{k_{A_{\zeta}}} a_{{\zeta},n} \cdot
\xi^n,\ \ \ldots,\ \ H_{\zeta} = \sum_{n = 0}^{k_{H_{\zeta}}} h_{{\zeta},n}
\cdot \xi^n.$$

The absence of magnets with multipole field expansion moments higher than the
quadrupole one implies that the horizontal trajectory position (direction) does
not depend on the vertical momentum component nor the vertical vertex
coordinate, and \textit{vice versa}. The best description of the scattered
proton transport for the ALFA detectors is given by:

\begin{equation}
\left.
\begin{array}{ccl}
  x & = & \sum\limits_{n = 0}^{k_{A_{x}}} a_{x,n} \cdot \xi ^n + \sum\limits_{n = 0}^{k_{B_{x}}} b_{x,n} \cdot \xi ^n\cdot x_{\mathrm{IP}} + \sum\limits_{n = 0}^{k_{E_{x}}} e_{x,n} \cdot \xi ^n\cdot x_{\mathrm{IP}}',  \\
  y & = & \sum\limits_{n = 0}^{k_{C_{y}}} c_{y,n} \cdot \xi ^n\cdot y_{\mathrm{IP}} + \sum\limits_{n = 0}^{k_{F_{y}}} f_{y,n} \cdot \xi ^n\cdot y_{\mathrm{IP}}',  \\
  x' & = & \sum\limits_{n = 0}^{k_{A_{x'}}} a_{x',n} \cdot \xi ^n + \sum\limits_{n = 0}^{k_{B_{x'}}} b_{x',n} \cdot \xi ^n\cdot x_{\mathrm{IP}} + \sum\limits_{n = 0}^{k_{E_{x'}}} e_{x',n} \cdot \xi ^n\cdot x_{\mathrm{IP}}',  \\
  y' & = & \sum\limits_{n = 0}^{k_{C_{y'}}} c_{y',n} \cdot \xi ^n\cdot y_{\mathrm{IP}} + \sum\limits_{n = 0}^{k_{F_{y'}}} f_{y',n} \cdot \xi ^n\cdot y_{\mathrm{IP}}',
\end{array}
\right\}
\label{eq:par3a}
\end{equation}

The exact values of the coefficients were found by fitting Eq. (\ref{eq:par1})
to the MAD-X PTC results. This parameterisation was validated with an
independent single diffractive event sample generated with \textsc{Pythia} 6.4
Monte Carlo~\cite{pythia}. In the simulation the interaction vertex position
was smeared appropriately for the discussed LHC tunes (see Tab.  \ref{tab_smearings}).
The parametrisation uncertainty was evaluated comparing the results obtained
with Eq.~(\ref{eq:par3a}) to those of the MAD-X PTC.  Results of this
comparison are presented in Table~\ref{tab_difference}. The differences between
the MAD-X PTC and the parametrisation results are much smaller than the
detector resolution (30 $\mu$m). This confirms that the parametrisation provides a good
description of the scattered proton trajectory positions and elevation angles at the
detectors positions.

\begin{table}[htbp]
	\caption{The parameterisation method uncertainty measured as the RMS of the
	difference between values given by MAD-X PTC and parameterisation
	equations for the early and nominal high $\beta^{*}$.}
	\label{tab_difference}	
	\begin{center}	
		\begin{tabular}{c c c c }
	   \hline \\ [-1.5ex]
		Variable & Unit & Early High $\beta^{*}$ & Nominal High $\beta^{*}$\\ [1ex] 
\hline \\[-1.5ex]
		$\Delta x$ & nm & 45.1 & 37.9 \\
		$\Delta y$ & nm & 50.4 & 22.9 \\
		$\Delta x'$ & nrad & 35.7 & 4.7 \\	
		$\Delta y'$ & nrad & 9.1 & 10.3  \\ [1ex]
\hline
		\end{tabular}
	\end{center}
\end{table}

\section{Unfolding Procedure}

The procedure of inferring the scattered proton momentum on the basis of the
detector measurements is called unfolding.  This can be done in various ways, here is is
performed by means of the minimisation of the following $\chi^2$ function:
\[ \chi^2(\mathbf{p}) = \frac{\left(x_1^D -
x_1(\mathbf{p})\right)^2}{\sigma^2_x} + \frac{\left(y_1^D -
y_1(\mathbf{p})\right)^2}{\sigma^2_y} + \frac{\left(x_2^D -
x_2(\mathbf{p})\right)^2}{\sigma^2_x} + \frac{\left(y_2^D -
y_2(\mathbf{p})\right)^2}{\sigma^2_y}, \]
where $(x_1^D,y_1^D)$ denote the coordinates of the scattered proton trajectory
position measured by the first station and $(x_1(\mathbf{p}), y_1(\mathbf{p}))$
-- the coordinates calculated using the transport parametrisation for  a proton
with momentum $\mathbf{p}$. The variables $x_2^D$, $y_2^D$, $x_2(\mathbf{p})$
and $y_2(\mathbf{p})$ refer to the positions at the second station. The
parametrisation does not describe correctly neither the losses of particles due
to the beam pipe nor the collimators apertures, but this drawback is of a minor
importance for solving the unfolding problem. 

\begin{figure}[p]
\centering
\vspace{-0.5cm}
\includegraphics[width=0.9\columnwidth]{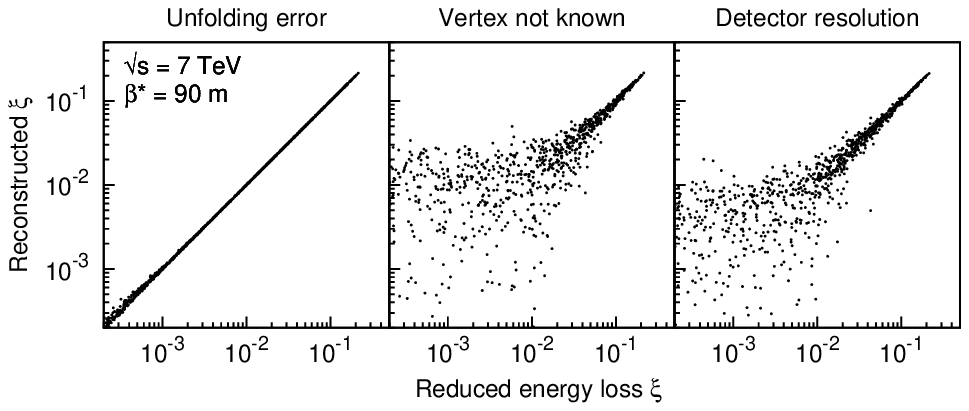}\\[2mm]
\includegraphics[width=0.9\columnwidth]{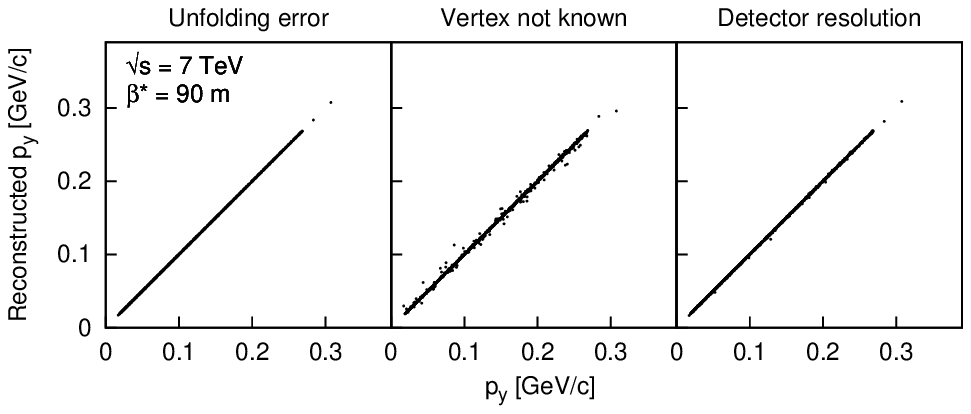}\\[2mm]
\includegraphics[width=0.9\columnwidth]{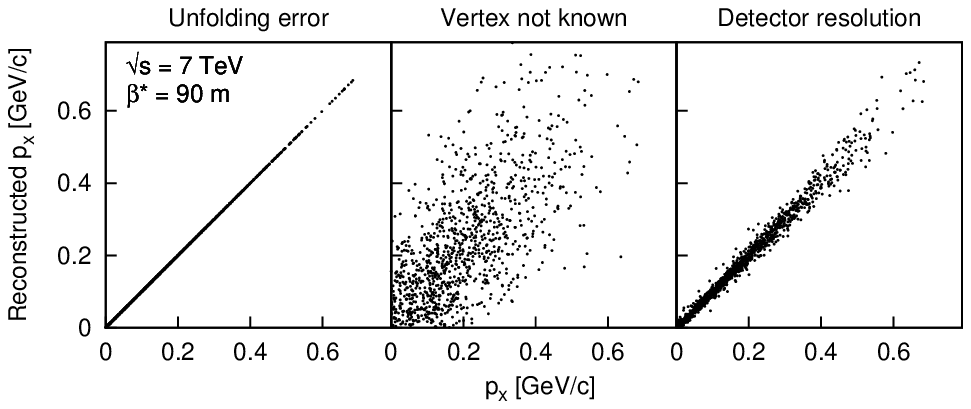}\\[2mm]
\includegraphics[width=0.9\columnwidth]{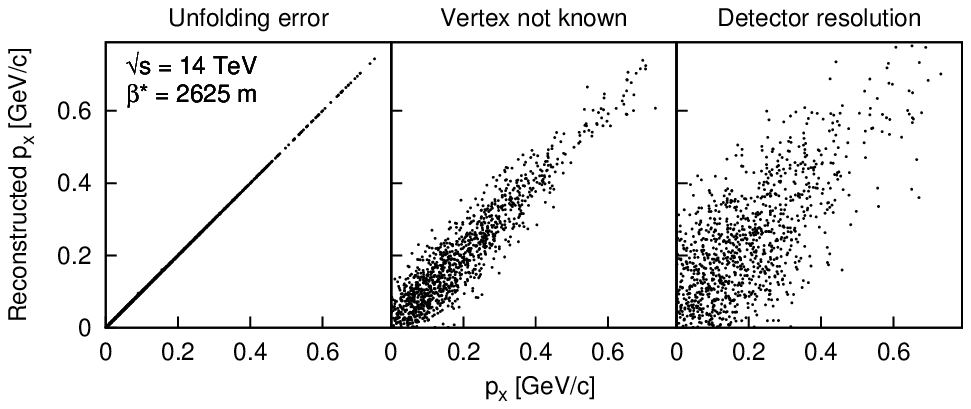}\\[2mm]
\caption{Correlations between the true and unfolded values of reduced energy loss and
transverse momentum components. The left column shows the ideal case where the
measurement is perfect, whereas the influence of experimental effects, the vertex smearing
and the detector resolution, are presented in the middle and right column,
respectively.}
\label{fig:correlation}
\end{figure}

\begin{figure}[ht]
  \begin{center}
    \includegraphics[width=0.49\columnwidth]{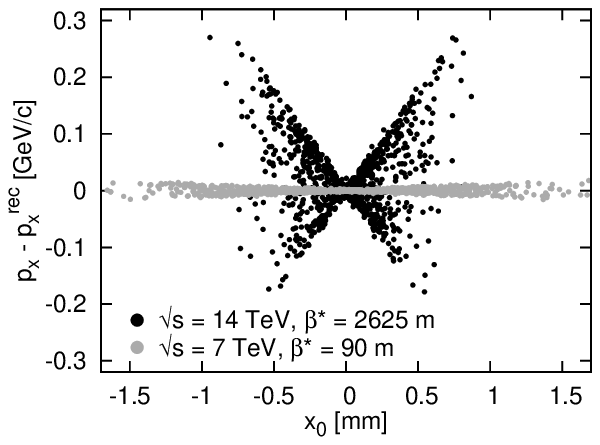}
    \hfill
    \includegraphics[width=0.49\columnwidth]{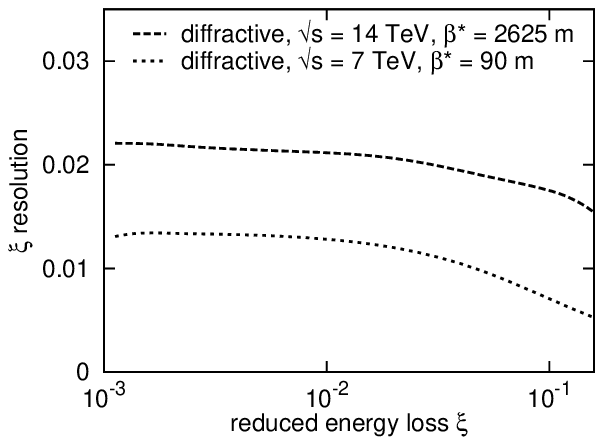}
  \end{center}
  \caption{Left: the correlation between the horizontal momentum reconstruction
  error and the horizontal vertex coordinate. Right: the reduced energy loss
  reconstruction resolution.}
  \label{fig:resolutions1}
\end{figure}

Results of the unfolding procedure are presented in Fig. \ref{fig:correlation}.
The correlations between the reconstructed and the generated value of the
reduced energy loss and the transverse momentum components of the scattered
proton are shown for the early high $\beta^{*}$ runs. For comparison the
horizontal transverse momentum unfolding results are added. This figure
presents also the influence of such experimental factors as the vertex smearing
and the detector spatial resolution ($\sigma_{x} = \sigma_y = 30\ \mu m$ is
assumed). The leftmost plots in Fig. \ref{fig:correlation} confirm the quality
of the parametrisation and the correctness of the unfolding procedure.  If
particular experimental effects are taken into account the unfolding accuracy
worsens. Due to the chosen optics mode, the momentum reconstruction works
better for the $y$ coordinate. The results show that in the early high
$\beta^{*}$ case the vertex smearing has a major impact on the horizontal
momentum reconstruction error whereas in the nominal high $\beta^*$ the leading
effect is governed by the detector resolution. The difference in the horizontal
momentum resolutions are due to the $x$-vertex spread. This is shown in Fig.
\ref{fig:resolutions1}, where a correlation between the $p_x$ unfolding error
and the horizontal vertex coordinate is plotted. It clearly shows that the
$x$-vertex spread has a much greater effect for the nominal high $\beta^*$
reconstruction accuracy.

\FloatBarrier

It is interesting to study what precision can be obtained for measurements of
the diffractively scattered protons. This would yield information on possible
measurements of diffraction that can be performed with the ALFA detectors. The
single diffractive dissociation events were generated with \textsc{Pythia}. The
final state forward protons were transported to the ALFA stations and the
actual positions of the trajectories were smeared according to the detector
spatial resolution. 

\begin{figure}[t]
  \begin{center}
    \includegraphics[width=0.49\textwidth]{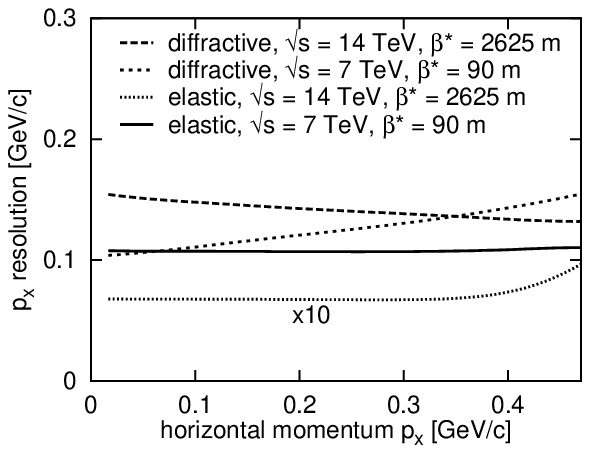}
    \hfill
    \includegraphics[width=0.49\textwidth]{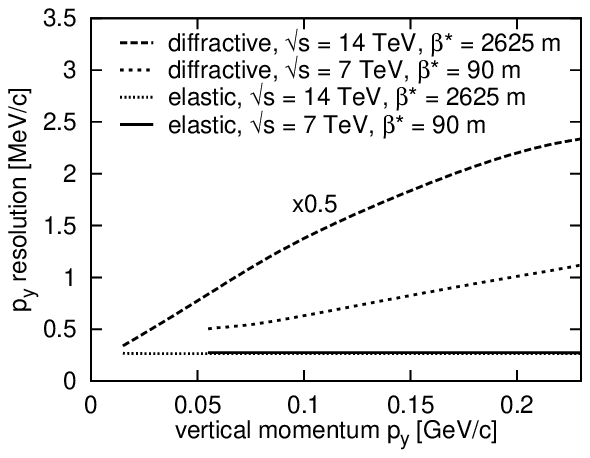}
  \end{center}
  \caption{The horizontal (left) and vertical (right) momentum resolution for
  elastically and diffractively scattered protons for early and nominal high
  $\beta^{*}$ optics.}
  \label{fig:resolutions2}
\end{figure}

Then, the unfolding procedure was performed and its errors were estimated by
fitting the distribution of the difference between the generated value and the
one obtained from the unfolding with the Gaussian distribution. These fits were
performed for different bins of $\xi$ (Fig. \ref{fig:resolutions1} right),
$p_x$ and $p_y$ (Fig. \ref{fig:resolutions2}). For comparison the momentum
resolutions were calculated for the elastically scattered protons. In this
case, the unfolding procedure was slightly different, because the $\chi^{2}$
minimisation was performed only in $p_{x}$ and $p_{y}$ variables as the proton energy is
fixed for such events.

One immediately notices that the early high $\beta^*$ offers better
opportunities for obtaining diffractively scattered proton kinematics. In that
case the energy loss reconstruction error is about 40 GeV which compares to
160~GeV for the nominal high $\beta^{*}$. One should notice large differences
between the resolutions in the horizontal and the vertical directions. This is
a direct consequence of the parallel-to-point focusing optics feature dedicated
to a precise measurement of small scattering angles in the vertical direction.

In the case of elastic scattering the situation is opposite: the horizontal
momentum reconstruction for the nominal high $\beta^*$ optics has much better
resolution than that for the early high $\beta^*$ one. The vertical momentum is
reconstructed very accurately and its reconstruction resolution is about 0.3 MeV/c
for both cases. This is because the nominal high $\beta^*$ optics is specially
designed for very precise elastic scattering measurements. However, one has to
remember that in the case of all high $\beta^{*}$ runs, the beam angular spread
can be a serious limiting factor (\textit{cf.} Tab. \ref{tab_smearings}). For
example, for the early high $\beta^*$ tune the beam transverse momentum spread
is about 20 times bigger than the $p_y$ reconstruction resolution. 

\section{Summary}

The transport of the elastically and diffractively scattered protons through
the LHC magnetic lattice for the high $\beta^{*}$ optics settings was
described.  The geometrical acceptance for the two most probable ALFA run
settings: the \textit{early high $\beta^{*}$} ($E_{\mathrm{beam}} =$ 3.5 TeV,
$\beta^{*} = $~90~m, $\epsilon^{*} = 2.5\ \mu$m$\cdot$rad) and the
\textit{nominal high $\beta^{*}$} ($E_{\mathrm{beam}} =$ 7 TeV, $\beta^{*} =$
2625~m, $\epsilon^{*} = $ 1 $\mu$m$\cdot$rad) was presented. Studies of the
transverse momentum acceptance as a function of the detector edge distance from
the beam centre reveals that the discussed $\beta^{*}$ settings are the most
desirable for the luminosity determination because the region of much lower
$p_T$ values can be accessed.

The transport parameterisation was introduced as the fastest simulation method
of the particle transport through the LHC magnetic lattice, which delivers an
analytical representation of the scattered proton position at the detector
stations. The accuracy of this method was shown to be much better than the
assumed spatial resolution of the detector.

Finally, the unfolding method was presented as a procedure to extract the
scattered proton energy and momentum at the Interaction Point from its
trajectory measurement at the forward detectors. The example of the ALFA case
shows that the proton energy can be reconstructed with precision of about 40
GeV in the case of early high $\beta^{*}$ and about 160 GeV for the second
discussed setting. The scattered proton momentum reconstruction precision is dominated by its
horizontal component resolution and is about $0.15$ GeV/c for both optics. In
the elastic scattering case the momentum resolution is still dominated by its
horizontal component reconstruction resolution, but the nominal high $\beta^*$
setting allows for twenty times better precision than the early high $\beta^{*}$
one.

\end{document}